\documentclass[
reprint,
amsmath,amssymb,
aps,
prl,
floatfix,
superscriptaddress
]{revtex4-1}

\usepackage{graphicx}					
\usepackage{dcolumn}					
\usepackage{bm}						
\usepackage{float}
\usepackage{gensymb}
\usepackage{multirow}

\newcommand{\KET}[1]{|#1\rangle}
\newcommand{\BRA}[1]{\langle#1|}

\renewcommand{\eqref}[1]{Eq.~(\ref{#1})}

\newcommand{\figref}[1]{Fig.~\ref{#1}}

\newcommand{\threevec}[1]{\mathbf{#1}}

\newcommand{\mhat}{\hat{\threevec{m}}}

\newcommand{\xhat}{\hat{\threevec{x}}}
\newcommand{\yhat}{\hat{\threevec{y}}}
\newcommand{\zhat}{\hat{\threevec{z}}}

\begin{document}

\preprint{APS/123-QED}


\title{Intrinsic spin currents in ferromagnets}


\author{V. P. Amin}
\email{vivek.amin@nist.gov}
\affiliation{
Maryland NanoCenter, University of Maryland, College Park, MD 20742
}
\affiliation{
Center for Nanoscale Science and Technology, National Institute of Standards and Technology, Gaithersburg, Maryland 20899, USA
}
\author{Junwen Li}
\affiliation{
Joint Institute for Computational Sciences, University of Tennessee, Knoxville, TN 37996
}
\author{M. D. Stiles}
\affiliation{
Center for Nanoscale Science and Technology, National Institute of Standards and Technology, Gaithersburg, Maryland 20899, USA
}
\author{P. M. Haney}
\affiliation{
Center for Nanoscale Science and Technology, National Institute of Standards and Technology, Gaithersburg, Maryland 20899, USA
}

\date{\today}


\begin{abstract}

First principles calculations show that electric fields applied to ferromagnets generate spin currents flowing perpendicularly to the electric field.  Reduced symmetry in these ferromagnets enables a wide variety of such spin currents.  However, the total spin current is approximately the sum of a magnetization-independent spin Hall current and an anisotropic spin anomalous Hall current.  Intrinsic spin currents are not subject to dephasing, enabling their spin polarizations to be misaligned with the magnetization, which enables the magnetization-independent spin Hall effect.  The spin Hall conductivity and spin anomalous Hall conductivities of transition metal ferromagnets are comparable to those found in heavy metals, opening new avenues for efficient spin current generation in spintronic devices.

\end{abstract}


\maketitle


\emph{Introduction---}Over the past few decades, the study of electrical spin current generation has focused on two systems: ferromagnets without spin-orbit coupling and nonmagnets with spin-orbit coupling.  In ferromagnets without spin-orbit coupling, real space and spin space are decoupled.  As a result, spin currents are simple products of particle flow and spin direction, each of which independently satisfies the system's symmetries.  Thus, carriers must flow parallel to the electric field and carrier spins must align with the magnetization.  In nonmagnets with spin-orbit coupling, real space and spin space are coupled.  Because of this coupling, the net spin current is no longer the simple product of particle flow and spin direction.  Any spin current that satisfies the system's symmetries is allowed.  Isotropic symmetry allows for spin currents in which the charge flow, spin flow, and spin direction are mutually orthogonal.  The generation of such spin currents satisfying these constraints is known as the spin Hall effect \cite{SHETheoryDyakonovPerel, SHETheoryHirsch, SHETheoryZhang, SHETheoryMurakami, SHETheorySinova, SHEExpKato, SHEExpWunderlich}.

Interest has now turned to spin currents generated in ferromagnets with spin-orbit coupling \cite{FSHEExpMiao,FSHEExpWu,FSHEExpGibbons,FSHEExpBose,FSHEExpDas,iSOCBaekAmin,FSHEExpTian,FSHEOmori}.  In these materials, the combination of spin-orbit coupling and the symmetries broken by the magnetization enable a wider array of spin currents than in nonmagnets.  Indeed, {\it all} symmetries are broken for a magnetization with arbitrary orientation with respect to the applied electric field.  As a result, ferromagnets with spin-orbit coupling exhibit richer spin current generation and are inevitably subject to greater confusion when interpreting experiments.

%

Despite this complexity, our central result shows that the magnetization-dependent spin current is described by two terms with familiar symmetry properties. The spin current is a tensor ${Q^\beta_\alpha}$ with two spatial indices: the subscript $\alpha$ specifies the flow direction and superscript $\beta$ specifies the spin direction. The spin current flowing in the $\xhat$-direction is a vector in spin space denoted by ${\bf Q}_x$.  Our findings can be conveniently expressed by taking the electric field along the $\yhat$-direction and considering the spin current flowing along the $\xhat$-direction.  The spin direction ${\bf Q}_x$ of this spin current depends on the magnetization direction $\threevec{m}$ as:
\begin{eqnarray}
{\bf Q}_x \approx (\sigma_\text{SHE} \zhat + \sigma_\text{SAHE} m_z \threevec{{\hat m}}) E.	\label{eq:MagDep}
\end{eqnarray}
The two terms in \eqref{eq:MagDep} have the symmetries of the spin Hall effect and the spin anomalous Hall effect, respectively. In what follows, we discuss how these effects contribute to spin current generation in ferromagnets.

\begin{figure}[b!]
	\centering
	\vspace{0pt}	
	\includegraphics[width=1\linewidth,trim={0.3cm 0.3cm 0.5cm 0.4cm},clip]{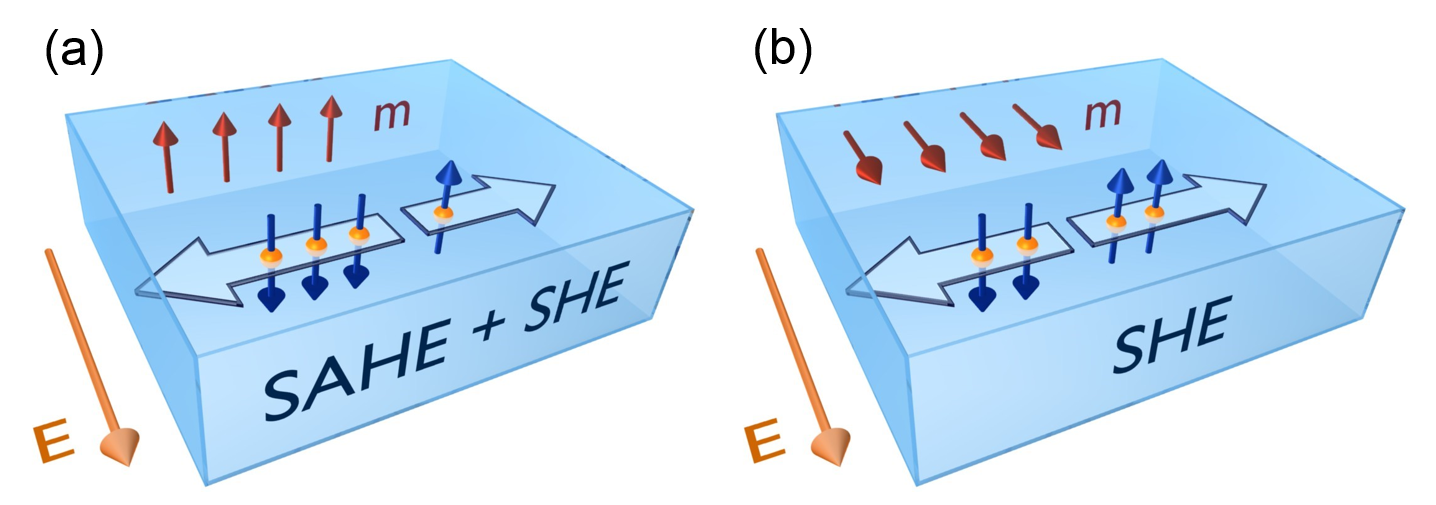}
	\caption{
	(Color online) Spin currents in ferromagnets with spin direction perpendicular to the flow and the electric field. (a) When the magnetization and spin direction are aligned, this spin current has contributions from both the spin Hall effect (SHE) and the spin anomalous Hall effect (SAHE). (b) When the magnetization and spin direction are perpendicular, the allowed spin current arises from the spin Hall effect alone.  
	}
	\vspace{0pt}
	\label{fig:SpinCurrentSchematics}
\end{figure}

The first term in \eqref{eq:MagDep} can be interpreted as a magnetization-independent spin Hall effect.  A counterintuitive feature of this spin current is that its spin direction may be misaligned with the magnetization.  Such spin currents violate the common assumption that spins misaligned with the magnetization rapidly precess in the magnetic exchange field and quickly dephase.  Thus, the presence of the first term in \eqref{eq:MagDep} suggests that the spin of eigenstates can be substantially misaligned with the magnetization in the presence of spin-orbit coupling \cite{Haney:2010}.

The second term in \eqref{eq:MagDep} is related to the anomalous Hall effect.  The anomalous Hall effect describes the current response perpendicular to an applied electric field $\threevec{E}$ in ferromagnets \cite{AHETheoryKundt,AHETheoryPughRostoker,AHETheoryKarplusLuttinger,AHEReviewNagosa}.  The anomalous Hall current flows along the $\mhat \times \threevec{E}$-direction. Since charge flow in ferromagnets is spin-polarized, the anomalous Hall current should be accompanied by a spin current with spin direction along $\mhat$.  The generation of such a spin current, called the spin anomalous Hall effect, was recently investigated theoretically \cite{AHEAMRTaniguchi}.  The second term in \eqref{eq:MagDep} describes a spin current with the same magnetization dependence as the spin anomalous Hall current.

Both the anomalous and spin Hall effects have extrinsic and intrinsic contributions.  Extrinsic contributions result from disorder scattering while intrinsic effects arise from the perturbation of electronic wave functions induced by the applied electric field.  It is now accepted that the intrinsic mechanism dominates the anomalous Hall response of typical transition metal ferromagnets \cite{AHEReviewNagosa}.  We therefore focus on this intrinsic regime.

In this work, we use first principles calculations to compute the intrinsic spin current conductivities for transition metal ferromagnets and show that that \eqref{eq:MagDep} describes the response.  The magnitude of the total spin current conductivity is substantial, suggesting that ferromagnets could be efficient and flexible generators of spin current.  We find that the spin direction of intrinsic spin currents is not subject to dephasing, which enables spin directions that are misaligned with the magnetization.  We also present a simple tight-binding model that captures the relevant physics which demonstrates why intrinsic spin currents are not subject to dephasing.

\begin{figure}[t!]
	\centering
	\vspace{0pt}	
	\includegraphics[width=1\linewidth,trim={0cm 0cm 0cm 0cm},clip]{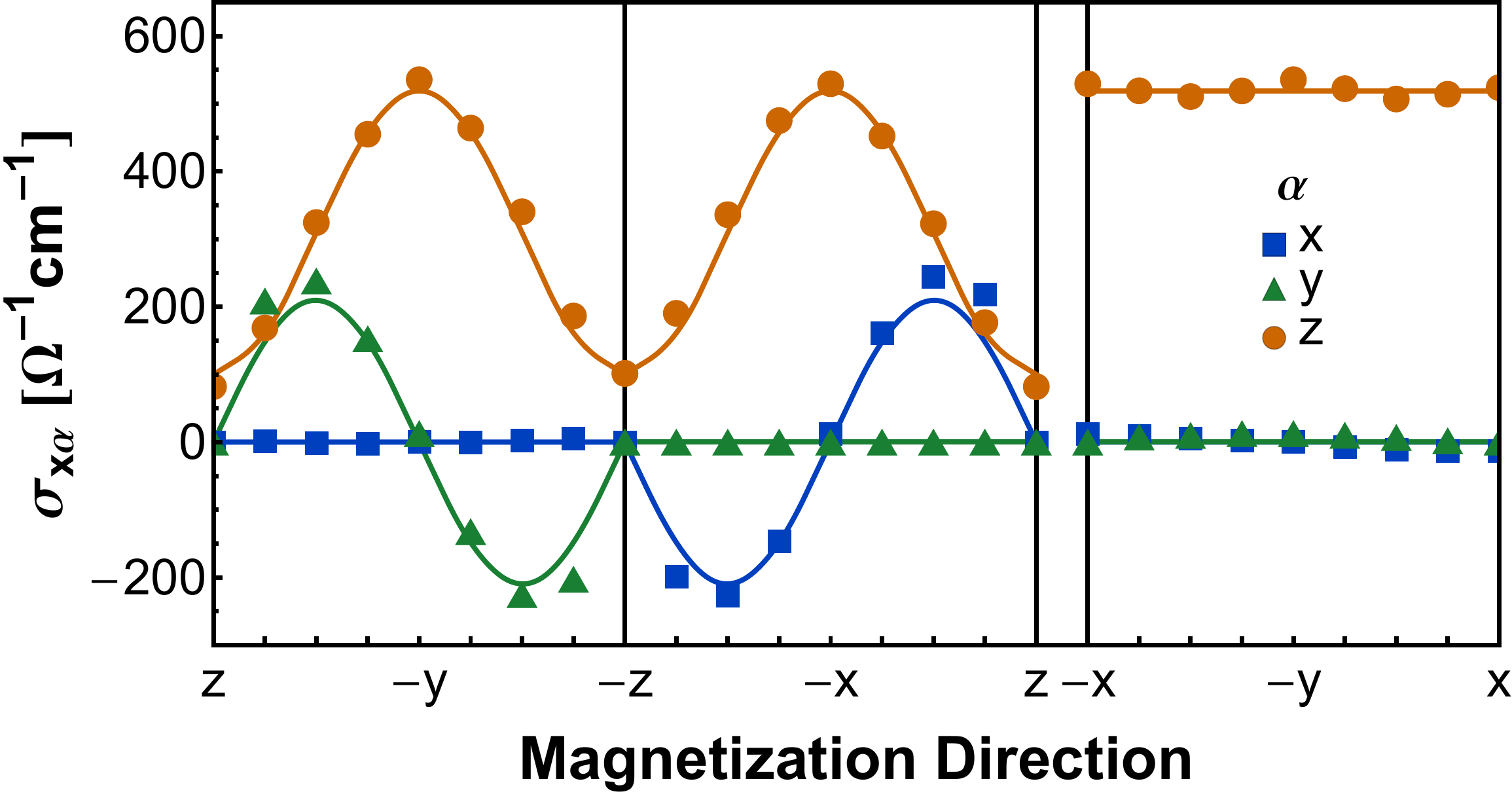}
	\caption{
	(Color online) Magnetization dependence of the $x$-flowing intrinsic currents generated by an applied electric field along $\yhat$ for BCC Fe.  The horizontal axis gives the magnetization direction, which is swept in the $z/y$, $z/x$, and $x/y$ planes.  Blue, green, and orange curves give spin currents with spin direction along $x$, $y$, and $z$ respectively.  Symbols denote values computed with density functional theory.  Solid curves show \eqref{eq:MagDep}, where the parameters $\sigma_\text{SHE}$ and $\sigma_\text{SAHE}$ are extracted from the first principles results as described in the text.
 	}
	\vspace{0pt}
	\label{fig:MagDep}
\end{figure}

\emph{First principles calculations---} We use the Kubo formalism to compute the spin current induced by an applied electric field.  For an electric field $E_\gamma$ in the $\gamma$-direction, the spin current conductivity tensor $\sigma$ yields the spin current according to $Q^\beta_{\alpha}=\sigma_{\alpha\gamma}^\beta E_\gamma$.  The expression for $\sigma$ in the clean limit for zero temperature is:
\begin{eqnarray}
\! \! \! \! \! \! \! \!
\sigma_{\alpha\gamma}^\beta = 
-\frac{2e^2}{\hbar} \, {\rm Im} \! \! \int \! \! \frac{d{\bf k}}{\left(2\pi\right)^3} 
\! \! \! \! \! \sum_{\substack{n\in {\rm occ.}\\m\in {\rm unocc.}}} \! \! \! \! \!
\frac{\langle\psi_m|{\hat Q}^\beta_\alpha|\psi_n\rangle\langle\psi_n|{\hat v}_{\gamma}|\psi_m\rangle}{\left(E_n-E_m\right)^2} \,
\label{eq:sigma}
\end{eqnarray}
where $v_\gamma= dH/dk_\gamma$ is the velocity operator along the $\gamma$-direction.  The spin current operator is given by ${\hat Q}^\beta_\alpha=\left(v_\alpha s_\beta + s_\beta v_\alpha\right)/2$, where $s_\beta$ is the $\beta$-component of the Pauli spin matrices.

\eqref{eq:sigma} is evaluated within density functional theory.  The ground state is computed with the {\sc Quantum ESPRESSO} package \cite{giannozzi2009quantum}, where we use the experimental lattice constants of (0.286, 0.352, 0.251)~${\rm nm}$ for Fe, Ni, and Co (hcp), respectively.  In each case, the plane-wave cutoff energy is set to $60~E_{\rm h}$, and a $12\times12\times12$ uniform $k$-point grid is used.  We use ultrasoft, fully relativistic pseudopotentials with a GGA functional.  For Ni, Ref.~\cite{fuh2011intrinsic} shows that the calculated anomalous Hall conductivity is closer to the experimental value using the GGA$+U$ method.  We therefore adopt a similar approach for Ni, presenting results for $U=1.9~{\rm eV}$ and $J=1.2~{\rm eV}$.  We find the final result is rather insensitive to this choice.

To evaluate \eqref{eq:MagDep} on a fine ${\bf k}$-point mesh, we performed Wannier interpolation using Wannier90 \cite{mostofi2008wannier90}. The Wannier projection is performed on the system with magnetization along the ${\bf {\hat z}}$-direction. To vary the magnetization direction, we decompose the Hamiltonian into components that are even and odd under time-reversal, and perform a spin-space rotation of the odd component to the desired orientation. The integral in \eqref{eq:sigma} is initially evaluated using a uniform mesh of $200^3$ ${\bf k}$-points.  We use an adaptive mesh procedure in which ${\bf k}$-dependent conductivity values exceeding $0.28~{\rm nm}^2$ are sampled on a refined mesh.  We continue mesh refinement until calculated values are converged to 1~\%.

\figref{fig:MagDep} shows the magnetization dependence of the intrinsic spin currents computed for Fe.  We again restrict our attention to spin currents flowing in the $\xhat$-direction generated by an electric field in the $\yhat$-direction. The phenomenological parameters $\sigma_{\rm SAHE}$ and $\sigma_{\rm SHE}$ are extracted from from the values of $\sigma^z_{xy}$ for the magnetization along $\zhat$ and $\yhat$.  The magnetization-dependence predicted \eqref{eq:MagDep} is shown in sold lines while the numerically computed values are shown as symbols. We find the full angular dependence is well-described by \eqref{eq:MagDep} \footnote{We find small deviations from the symmetry-required form of the conductivity.  This is due to the fact that the Wannier projection results in a Hamiltonian which does not exactly respect the symmetries of the crystal lattice.  We have also constructed symmetrized Hamiltonians, and find that symmetry-related errors introduced by the Wannier process are small, on the order of $(0.01~,0.1,~10)~\left({\rm \Omega\cdot cm}\right)^{-1}$ for Fe, Ni, and Co, respectively.}.

The values for $\sigma_\text{AHE}$, $\sigma_\text{SHE}$, and $\sigma_\text{SAHE}$ obtained for Fe, Co, and Ni are shown in Table \ref{table:FP}.  The magnitude of the spin Hall conductivity is substantial and indicates the potential for ferromagnets to be flexible and effective sources of spin current.  The response of the cubic crystals Fe and Ni coincide well with \eqref{eq:MagDep}, while HCP Co exhibits substantially more anisotropy arising from the crystal anisotropy.

\begin{table}[t!]
\setlength{\tabcolsep}{6pt}
    \begin{tabular}{|c||c|c|c|c|}
    \hline
                     						& $\sigma_{{\rm AHE}}$	& $\sigma_{{\rm SHE}}$	& $\sigma_{{\rm SAHE}}$	\\ \hline
Fe									& 720 					& 519 					& -419 					\\ \hline
Ni									& -1326 					& 1688 					& -728 					\\ \hline
Co $(\threevec{E}~||~\threevec{a})$		& 454 					& -130 					& -8 					\\ \hline
Co $(\threevec{E}~||~\threevec{c})$		& 159 					& 1074 					& -1004 					\\ \hline
    \end{tabular}
    \caption{Computed conductivity components (units of $\Omega^{-1} \text{cm}^{-1}$).  For Co, results are shown for the electric field $\threevec{E}$ along $\threevec{a}$ and $\threevec{c}$ to show the crystal-induced anisotropy.}
    \label{table:FP}
\end{table}

\begin{figure}[b!]
	\centering
	\vspace{0pt}	
	\includegraphics[width=1.0\linewidth,trim={0cm 0cm 0cm 1cm},clip]{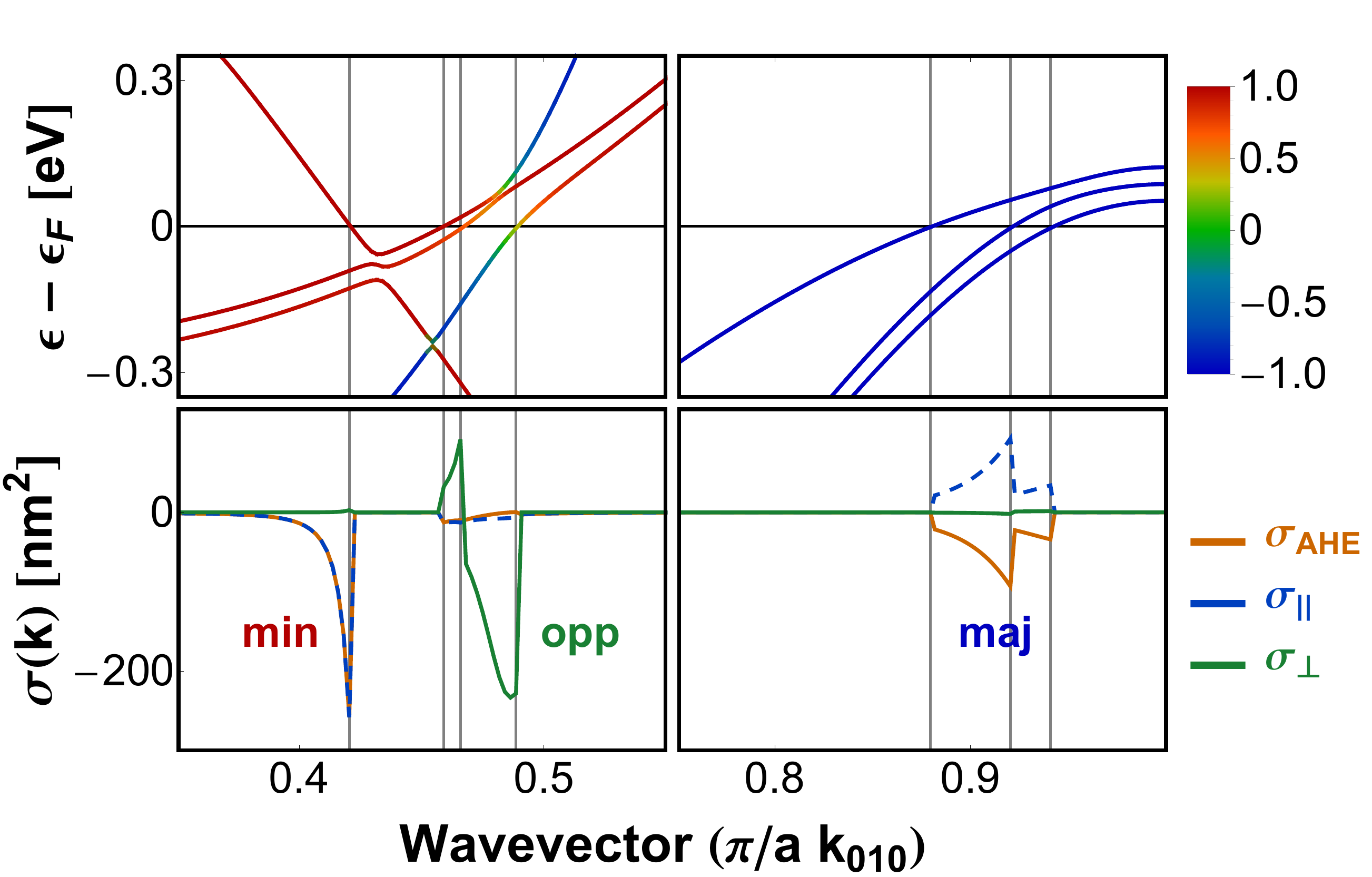}
	\caption{
	(Color online) Band structure near the Fermi energy $\epsilon_F$  (top) and $k$-dependent intrinsic conductivities (bottom) for BCC Fe, where $\mhat = \left(\yhat + \zhat\right)/\sqrt{2}$.  Band color gives value of $\bf{s}\cdot \bf{m}$, with blue (red) bands corresponding to majority (minority) carriers.  Avoided crossings between like (opposite) spin bands contribute strongest to $\sigma_\parallel$ ($\sigma_\perp$), which describes the spin current with spin direction parallel (perpendicular) to $\mhat$.
 	}
	\vspace{0pt}
	\label{fig:BandCrossing}
\end{figure}


To analyze the microscopic origin of our results, it is useful to write the spin current of \eqref{eq:MagDep} in terms of spin components parallel and perpendicular to the magnetization.  For an electric field in the $\yhat$-direction and flow in the $\xhat$-direction, the appropriate form is given below \footnote{The general form for the spin current tensor for flow in the ${\bf {\hat n}}$-direction, spin-polarized in the ${\bf {\hat s}}$-direction under an applied field ${\bf E}$ is: $Q^\threevec{\hat{n}}_\threevec{\hat{s}} = \sigma_{\parallel}\left({{\bf \hat s}} \cdot {{\bf \hat m}}\right) | {{\bf \hat m}} \cdot \left({\bf E}\times {{\bf \hat n}}\right)| + \sigma_{\perp} {{\bf \hat s}} \cdot \left( {{\bf \hat m}} \times \left( {{\bf \hat m}} \times \left( {\bf E} \times {{\bf \hat n}} \right) \right) \right)$}.
\begin{eqnarray}
{\bf Q}_x = \sigma_\parallel m_z \mhat + \sigma_\perp \left(\mhat \times \mhat \times \zhat\right). \label{eq:Qys2}
\end{eqnarray}
where we refer to $\sigma_{\parallel}$ ($\sigma_{\perp}$) as the longitudinal (transverse) spin Hall conductivity.  Note that $\sigma_\parallel=\sigma_{\rm SAHE} + \sigma_{\rm SHE}$ and $\sigma_\perp=\sigma_{\rm SHE}$.

In \figref{fig:BandCrossing}, we show the band structure and the $k$-resolved conductivities for Fe with magnetization along $\mhat=\left(\yhat+\zhat\right)/\sqrt{2}$.  As expected, contributions to conductivities exhibit peaks at avoided band crossings near the Fermi energy.  Peaks in the anomalous Hall conductivity $\sigma_{\rm AHE}$ and the longitudinal spin Hall conductivity $\sigma_\parallel$ (denoted by ``maj'' and ``min'') can be associated with interband coupling between states with the same spin.  The magnetization-aligned spin current for these states is approximately equal to the charge current up to a sign determined by the spin direction.

We also observe a peak in $\sigma_\perp$ (denoted by ``opp'') that arises from coupling between bands with opposite spin, suggesting that different components of the conductivity can be associated with different types of band pairs.  To quantify this association, we partition the sum over bands $n$ and $m$ in \eqref{eq:sigma} into two parts according to the sign of $P=\left({\bf s}_n \cdot {\bf s}_m\right)$, {\it i.e} pairs of like-spin bands ($P>0$) versus pairs of opposite-spin bands ($P<0$).  We find that 95~\% of the magnitude of $\sigma_{\rm AHE}$ comes from spin-like band pairs \footnote{Ref. \cite{wang2006ab} partitions contributions to $\sigma_{\rm AHE}$ according to band-pair type and finds a different values.  However the sum is partitioned differently in the two cases: Ref. \cite{wang2006ab} work assigns one band-pair type to each ${\bf k}$-point, whereas we track all band-pair types.} while 96~\% of the magnitude of $\sigma_\perp$ comes from spin-opposite band pairs.  As we discuss in the next section, a wave function which carries a spin current with spin direction transverse to the magnetization requires a superposition of majority and minority spin states.  It is therefore not surprising that $\sigma_\perp$ arises from spin-opposite band pairs.  The longitudinal spin Hall conductivity $\sigma_\parallel$ has contributions from both band pair types: 63~\% arise from spin-like pairs and 36~\% arise from spin-opposite pairs.

We note that the longitudinal spin Hall conductivity $\sigma_\parallel$ is not given by the product of $\sigma_{\rm AHE}$ with the material's bulk spin polarization.  The reason for this difference is that the bulk spin polarization is determined by factors such as the relative number of majority and minority spin states at the Fermi energy and their scattering times, while the intrinsic spin Hall conductivity is determined by other factors such as the relative number of spin-opposite versus spin-like band pairs straddling the Fermi energy.

\begin{figure}[t!]
	\centering
	\vspace{0pt}	
	\includegraphics[width=1\linewidth,trim={1cm 2cm 1cm 0cm},clip]{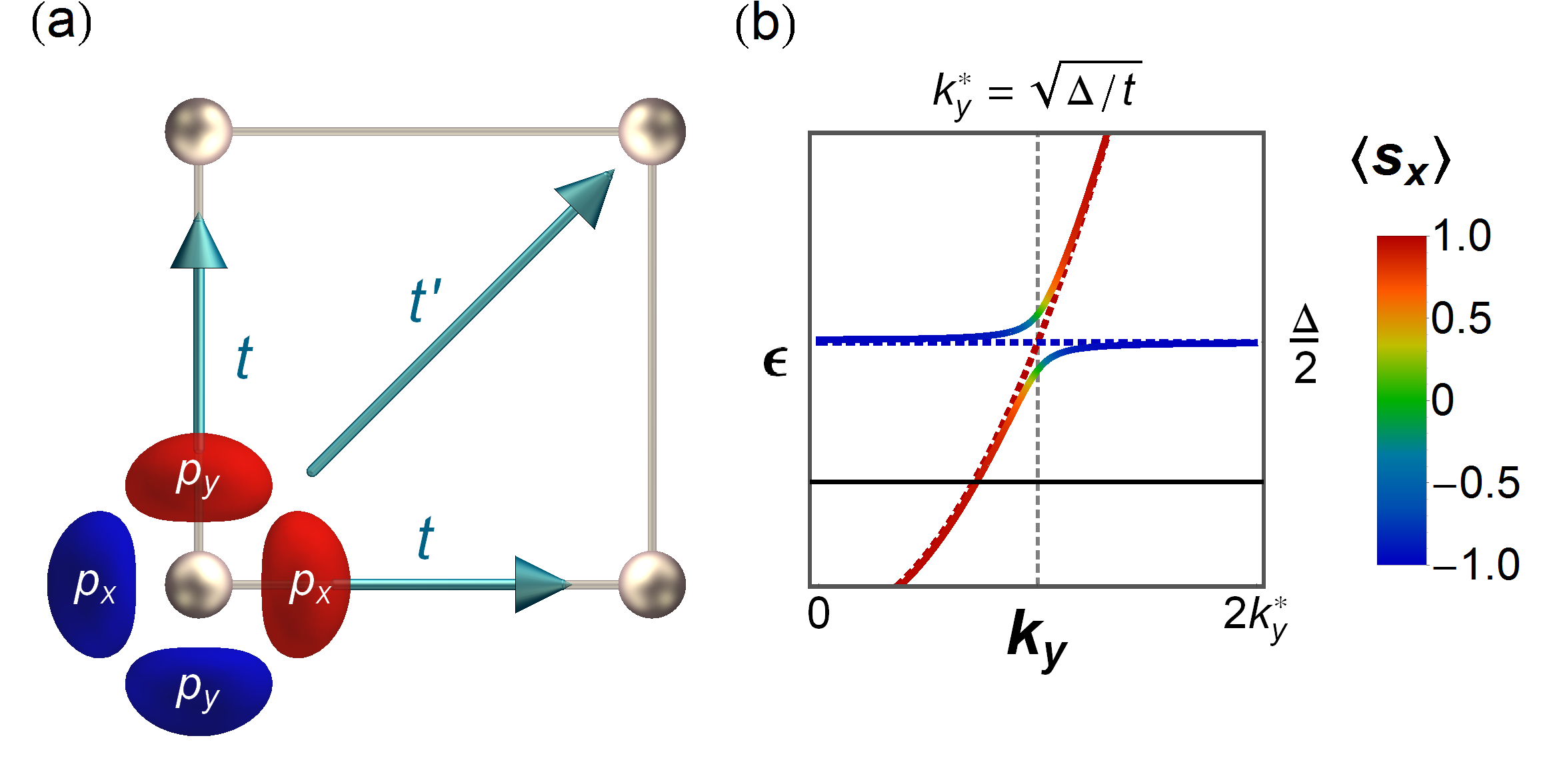}
	\caption{
	(Color online) (a) Schematic of the tight binding model, which describes electrons occupying $p$-orbitals on a 2D square lattice with nearest ($t$) and next-nearest ($t'$) neighbor hopping ($\mhat = \xhat$).  (b) Plot of the avoided band crossing along the $k_y$ axis.  The spin density vanishes at the avoided crossing between majority and minority bands.  However, the spin current with flow and spin direction transverse to the magnetization ($Q^x_z$) does not vanish at the avoided crossing.
	}
	\vspace{0pt}
	\label{fig:TBModel}
\end{figure}

\emph{Toy model for intrinsic spin Hall conductivity ---} We use a simple tight-binding model to demonstrate an intrinsic spin current with spin direction transverse to the magnetization.  The model consists of a 2-d square lattice in the $x/y$ plane with $p_x$ and $p_y$ orbitals.  We include nearest neighbor hopping $t$ and next-nearest-neighbor hopping $t'$ (see \figref{fig:TBModel}a).  Both types of hopping are required for this model to yield a spin Hall current with spins transverse to the magnetization.  The magnetization direction is along $\xhat$ and leads to a spin-dependent exchange splitting $\Delta$. We express the Hamiltonian $H$ in terms of the outer product of orbital space $(p_x,p_y)$ and spin space $(\uparrow,\downarrow)$, therefore making $H$ a $4\times 4$ matrix.  We can concisely write $H$ with the Bloch factor $e^{i\bm{k}\cdot\bm{r}}$ absorbed as:
\begin{align}
\! H =	\begin{pmatrix}
		t k_x^2		&	t'k_x k_y		\\
		t' k_x k_y 	&	t k_y^2		\\
		\end{pmatrix}
		\otimes I_s + \Delta I_p \otimes s_x + \lambda L_z\otimes s_z, \label{eq:H}
\end{align}
where $I_s$ and $I_p$ are identity operators in spin and orbital space, respectively.  The first term of \eqref{eq:H} describes hopping in the $(p_x,p_y)$ basis.  The hopping is spin-independent, so the first term consists of two copies of the orbital-dependent hopping matrix along the spin-diagonal.  Note that the Bloch wave vector ${\bf k}$ is dimensionless (scaled by the inverse lattice constant $1/a$) and that we take the small ${\bf k}$ limit.  The second term gives the orbital-independent magnetic exchange splitting for magnetization direction along $\xhat$.  The third term captures atomic spin-orbit coupling.  The full form of spin-orbit coupling is ${\bf L}\cdot {\bf s}$, however Ref. \cite{SHETheoryTanaka} shows that the $s_z L_z$ term contributes most substantially to the spin Hall conductivity.  We therefore include only this term for simplicity.  The results do not change appreciably if we use the full spin orbit coupling form ${\bf L}\cdot {\bf s}$.

To develop the simplest demonstration of a spin Hall current with spins transverse to the magnetization, we begin by considering the wave functions for $k_x=0$ with no spin-orbit coupling ($\lambda=0$). The eigenstates are pure $p_x$ or $p_y$ orbitals with spins along the $x$ direction.
%
The addition of spin-orbit coupling modifies the eigenstates, and its effect is strongest near degeneracies.  For the degenerate point $k_y^*$ shown in \figref{fig:TBModel}(b), spin-orbit coupling splits the states according to their total angular momentum. Recall that the total angular momentum eigenstates are given by:
\begin{eqnarray}
\KET{J_{\pm1/2}} &=& \big{(} \KET{p_x} \pm i\KET{p_y} \big{)} \otimes \KET{\! \downarrow \! (\uparrow) \,} \\
\KET{J_{\pm3/2}} &=& \big{(} \KET{p_x} \pm i\KET{p_y} \big{)} \otimes \KET{\! \uparrow \! (\downarrow) \,}
\end{eqnarray}
where the arrow in parenthesis of the spin ket is paired with the lower sign.  Excluding the two eigenstates far away from the avoided crossing, the eigenstates for the spin-orbit coupled system at $k_y^*$ are:
\begin{eqnarray}
\KET{\Psi_1} & =& \KET{J_{+1/2}} - \KET{J_{-1/2}} \\
\KET{\Psi_2} & =& \KET{J_{+3/2}} - \KET{J_{-3/2}} .
\end{eqnarray}
The spin expectation values for $\KET{\Psi_1}$ and $\KET{\Psi_2}$ vanish due to mixing of the majority and minority states. Similar behavior is also observed in the band structure for Fe at crossings of spin-opposite bands (see \figref{fig:BandCrossing}).

The application of an electric field $\threevec{E}$ along $\yhat$ induces interband coupling with amplitude $iqEa\BRA{\Psi_1} \partial H/\partial k_y \KET{\Psi_2}/(\epsilon_1-\epsilon_2)^2$, where $q$ is the magnitude of electron charge and $\epsilon_{1,2}$ is the energy of $\KET{\Psi_{1,2}}$.  The perturbed wave function for $\KET{\Psi_1'}$ therefore reads:
\begin{eqnarray}
\KET{\Psi_1'} = \KET{\Psi_1} + iqEa\frac{k_y^* t}{\lambda^2} \KET{\Psi_2} \label{eq:wfE}
\end{eqnarray}
where $\lambda$ is the energy splitting induced by spin-orbit coupling. Evaluating the expectation value of the transverse spin current $Q^x_z$ with the perturbed wave function leads to the following result (to lowest order in $E$):
\begin{eqnarray}
\BRA{\Psi_1'} \hat{Q}^x_z \KET{\Psi_1'} = -\frac{qEa}{\hbar}\frac{t t'\left(k_y^*\right)^2}{\lambda} = -\frac{qEa}{\hbar}\frac{t' \Delta}{\lambda^2} \label{eq:Qtoy}
\end{eqnarray}
Note that the $\left(k_y^*\right)^2$ factor implies that contributions from $+k_y^*$ and $-k_y^*$ do not cancel.  The second equality in \eqref{eq:Qtoy} follows from the expression for $k_y^*$ given in \figref{fig:TBModel}.  Although the value of the Hall current at $k_y^*$ diverges as $\lambda \rightarrow 0$, the total Hall current conductivity integrated over $k$ goes continuously to zero as $\lambda \rightarrow 0$.

It is instructive to rewrite the wave function of \eqref{eq:wfE} in the $(p_x,p_y)\otimes(\uparrow,\downarrow)$ basis:
\begin{eqnarray}
\KET{\Psi_1'}		&=		&\big{(} \KET{p_x} - (i - E')\KET{p_y} \big{)} \otimes \KET{\! \uparrow \,}			\nonumber \\
				&&-~	 \big{(} \KET{p_x} + (i - E')\KET{p_y} \big{)} \otimes \KET{\! \downarrow \,}		\label{eq:wf2}
\end{eqnarray}
where $E'=2qEa k_y^* t/\lambda^2$. It is straightforward to show that, to linear order in $E$, this wave function has vanishing spin density while carrying a spin current given by \eqref{eq:Qtoy}. This example illustrates that spin currents are inaccurately described as ``spin densities that flow''.  That is, in the presence of spin orbit-coupling, spin currents cannot be treated as direct products of a flow direction and a spin direction.

\eqref{eq:Qtoy} shows that the spin Hall conductivity requires next-nearest neighbor hopping $t'$.  This requirement was discussed in Refs. \cite{SHETheoryTanaka,kontani2007intrinsic}, which emphasize the importance of interorbital hopping to the formation of Hall currents.  In our model, the transverse spin current along $\xhat$ arises from ``diagonal'' hopping between orbitals $\KET{p_x}$ and $\KET{p_y}$ that is asymmetric in the $+\xhat$ and $-\xhat$ directions.  The perturbing term $E'$ in \eqref{eq:wf2} enables this type of hopping, which can be understood as a consequence of interference between the orbital moment and the Bloch wave \cite{RasbhaTheoryPark}.


\emph{Discussion and conclusion---}Intrinsic spin currents with spin direction transverse to the magnetization do not dephase.  These spin currents are protected from dephasing because they are carried by perturbed eigenstates that superimpose different spin states with the same Bloch wavevector.  As these perturbed eigenstates propagate in space, the two spin components do not accumulate any relative phase, and hence do not precess and subsequently dephase.  However, dephasing could occur via spin-dependent scattering at interfaces, resulting in spin torques.  This suggests that spin-orbit torque must be reexamined in magnetic heterostructures to account for spin currents generated by ferromagnetic layers.  For extrinsic mechanisms, where the scattering site breaks crystal translation symmetry, the scattered wave functions with moments transverse to the magnetization are coherent superpositions of states with different wave vectors.  The different wave vectors lead to rapid precession and dephasing.  Thus, we expect that extrinsic contributions will be strongest for spin currents aligned with the magnetization. We leave this topic for future work.

Spin currents in ferromagnets have been measured in several experiments. Some isolate the spin Hall current whose spin direction is transverse to the magnetization \cite{iSOCBaekAmin,FSHEExpTian}. Others measure the spin current with a magnetization-aligned spin direction \cite{FSHEExpMiao,FSHEExpWu,FSHEExpGibbons,FSHEExpBose,FSHEOmori}, as would be expected for bulk spin currents in ferromagnets that have spin directions aligned with the magnetization \cite{AHEAMRTaniguchi}.  Das et al. quantify contributions from both transverse and magnetization-aligned spin directions within Py \cite{FSHEExpDas}, providing experimental evidence supporting some of the results presented here.

In this work, we demonstrated that spin currents in ferromagnets generated by the intrinsic mechanism are well approximated as the sum of a spin Hall current and a spin anomalous Hall current.  The spin Hall current has a spin direction transverse to the magnetization and could generate spin-orbit torques at material interfaces.  In transition metal ferromagnets, we find that these spin currents are comparable in magnitude to those generated in heavy metals.  This work should have immediate bearing on recent experiments which probe the spin anomalous Hall and spin Hall effects in ferromagnets and on investigations of spin-orbit torque in magnetic heterostructures.

\begin{acknowledgments}
The authors thank Jairo Sinova for useful conversations and Robert McMichael and Xin Fan for critical readings of the manuscript.  VA acknowledges support under the Cooperative Research Agreement between the University of Maryland and the National Institute of Standards and Technology Center for Nanoscale Science and Technology, Award 70NANB14H209, through the University of Maryland.
\end{acknowledgments}


\bibliography{FMSHEbib}

\providecommand{\noopsort}[1]{}\providecommand{\singleletter}[1]{#1}%
\begin{thebibliography}{31}%
\makeatletter
\providecommand \@ifxundefined [1]{%
 \@ifx{#1\undefined}
}%
\providecommand \@ifnum [1]{%
 \ifnum #1\expandafter \@firstoftwo
 \else \expandafter \@secondoftwo
 \fi
}%
\providecommand \@ifx [1]{%
 \ifx #1\expandafter \@firstoftwo
 \else \expandafter \@secondoftwo
 \fi
}%
\providecommand \natexlab [1]{#1}%
\providecommand \enquote  [1]{``#1''}%
\providecommand \bibnamefont  [1]{#1}%
\providecommand \bibfnamefont [1]{#1}%
\providecommand \citenamefont [1]{#1}%
\providecommand \href@noop [0]{\@secondoftwo}%
\providecommand \href [0]{\begingroup \@sanitize@url \@href}%
\providecommand \@href[1]{\@@startlink{#1}\@@href}%
\providecommand \@@href[1]{\endgroup#1\@@endlink}%
\providecommand \@sanitize@url [0]{\catcode `\\12\catcode `\$12\catcode
  `\&12\catcode `\#12\catcode `\^12\catcode `\_12\catcode `\%12\relax}%
\providecommand \@@startlink[1]{}%
\providecommand \@@endlink[0]{}%
\providecommand \url  [0]{\begingroup\@sanitize@url \@url }%
\providecommand \@url [1]{\endgroup\@href {#1}{\urlprefix }}%
\providecommand \urlprefix  [0]{URL }%
\providecommand \Eprint [0]{\href }%
\providecommand \doibase [0]{http://dx.doi.org/}%
\providecommand \selectlanguage [0]{\@gobble}%
\providecommand \bibinfo  [0]{\@secondoftwo}%
\providecommand \bibfield  [0]{\@secondoftwo}%
\providecommand \translation [1]{[#1]}%
\providecommand \BibitemOpen [0]{}%
\providecommand \bibitemStop [0]{}%
\providecommand \bibitemNoStop [0]{.\EOS\space}%
\providecommand \EOS [0]{\spacefactor3000\relax}%
\providecommand \BibitemShut  [1]{\csname bibitem#1\endcsname}%
\let\auto@bib@innerbib\@empty
\bibitem [{\citenamefont {D'yakonov}\ and\ \citenamefont
  {Perel}(1971)}]{SHETheoryDyakonovPerel}%
  \BibitemOpen
  \bibfield  {author} {\bibinfo {author} {\bibfnamefont {M.~I.}\ \bibnamefont
  {D'yakonov}}\ and\ \bibinfo {author} {\bibfnamefont {V.~I.}\ \bibnamefont
  {Perel}},\ }\href@noop {} {\ \textbf {\bibinfo {volume} {13}},\ \bibinfo
  {pages} {467} (\bibinfo {year} {1971})}\BibitemShut {NoStop}%
\bibitem [{\citenamefont {Hirsch}(1999)}]{SHETheoryHirsch}%
  \BibitemOpen
  \bibfield  {author} {\bibinfo {author} {\bibfnamefont {J.~E.}\ \bibnamefont
  {Hirsch}},\ }\href {\doibase 10.1103/PhysRevLett.83.1834} {\bibfield
  {journal} {\bibinfo  {journal} {Phys. Rev. Lett.}\ }\textbf {\bibinfo
  {volume} {83}},\ \bibinfo {pages} {1834} (\bibinfo {year}
  {1999})}\BibitemShut {NoStop}%
\bibitem [{\citenamefont {Zhang}(2000)}]{SHETheoryZhang}%
  \BibitemOpen
  \bibfield  {author} {\bibinfo {author} {\bibfnamefont {S.}~\bibnamefont
  {Zhang}},\ }\href {\doibase 10.1103/PhysRevLett.85.393} {\bibfield  {journal}
  {\bibinfo  {journal} {Phys. Rev. Lett.}\ }\textbf {\bibinfo {volume} {85}},\
  \bibinfo {pages} {393} (\bibinfo {year} {2000})}\BibitemShut {NoStop}%
\bibitem [{\citenamefont {Murakami}\ \emph {et~al.}(2003)\citenamefont
  {Murakami}, \citenamefont {Nagaosa},\ and\ \citenamefont
  {Zhang}}]{SHETheoryMurakami}%
  \BibitemOpen
  \bibfield  {author} {\bibinfo {author} {\bibfnamefont {S.}~\bibnamefont
  {Murakami}}, \bibinfo {author} {\bibfnamefont {N.}~\bibnamefont {Nagaosa}}, \
  and\ \bibinfo {author} {\bibfnamefont {S.-C.}\ \bibnamefont {Zhang}},\ }\href
  {\doibase http://dx.doi.org/10.1126/science.1087128} {\bibfield  {journal}
  {\bibinfo  {journal} {Science}\ }\textbf {\bibinfo {volume} {301}},\ \bibinfo
  {pages} {1348} (\bibinfo {year} {2003})}\BibitemShut {NoStop}%
\bibitem [{\citenamefont {Sinova}\ \emph {et~al.}(2004)\citenamefont {Sinova},
  \citenamefont {Culcer}, \citenamefont {Niu}, \citenamefont {Sinitsyn},
  \citenamefont {Jungwirth},\ and\ \citenamefont
  {MacDonald}}]{SHETheorySinova}%
  \BibitemOpen
  \bibfield  {author} {\bibinfo {author} {\bibfnamefont {J.}~\bibnamefont
  {Sinova}}, \bibinfo {author} {\bibfnamefont {D.}~\bibnamefont {Culcer}},
  \bibinfo {author} {\bibfnamefont {Q.}~\bibnamefont {Niu}}, \bibinfo {author}
  {\bibfnamefont {N.~A.}\ \bibnamefont {Sinitsyn}}, \bibinfo {author}
  {\bibfnamefont {T.}~\bibnamefont {Jungwirth}}, \ and\ \bibinfo {author}
  {\bibfnamefont {A.~H.}\ \bibnamefont {MacDonald}},\ }\href {\doibase
  10.1103/PhysRevLett.92.126603} {\bibfield  {journal} {\bibinfo  {journal}
  {Phys. Rev. Lett.}\ }\textbf {\bibinfo {volume} {92}},\ \bibinfo {pages}
  {126603} (\bibinfo {year} {2004})}\BibitemShut {NoStop}%
\bibitem [{\citenamefont {Kato}\ \emph {et~al.}(2004)\citenamefont {Kato},
  \citenamefont {Myers}, \citenamefont {Gossard},\ and\ \citenamefont
  {Awschalom}}]{SHEExpKato}%
  \BibitemOpen
  \bibfield  {author} {\bibinfo {author} {\bibfnamefont {Y.~K.}\ \bibnamefont
  {Kato}}, \bibinfo {author} {\bibfnamefont {R.~C.}\ \bibnamefont {Myers}},
  \bibinfo {author} {\bibfnamefont {A.~C.}\ \bibnamefont {Gossard}}, \ and\
  \bibinfo {author} {\bibfnamefont {D.~D.}\ \bibnamefont {Awschalom}},\ }\href
  {\doibase http://dx.doi.org/10.1126/science.1105514} {\bibfield  {journal}
  {\bibinfo  {journal} {Science}\ }\textbf {\bibinfo {volume} {306}},\ \bibinfo
  {pages} {1910} (\bibinfo {year} {2004})}\BibitemShut {NoStop}%
\bibitem [{\citenamefont {Wunderlich}\ \emph {et~al.}(2005)\citenamefont
  {Wunderlich}, \citenamefont {Kaestner}, \citenamefont {Sinova},\ and\
  \citenamefont {Jungwirth}}]{SHEExpWunderlich}%
  \BibitemOpen
  \bibfield  {author} {\bibinfo {author} {\bibfnamefont {J.}~\bibnamefont
  {Wunderlich}}, \bibinfo {author} {\bibfnamefont {B.}~\bibnamefont
  {Kaestner}}, \bibinfo {author} {\bibfnamefont {J.}~\bibnamefont {Sinova}}, \
  and\ \bibinfo {author} {\bibfnamefont {T.}~\bibnamefont {Jungwirth}},\ }\href
  {\doibase 10.1103/PhysRevLett.94.047204} {\bibfield  {journal} {\bibinfo
  {journal} {Phys. Rev. Lett.}\ }\textbf {\bibinfo {volume} {94}},\ \bibinfo
  {pages} {047204} (\bibinfo {year} {2005})}\BibitemShut {NoStop}%
\bibitem [{\citenamefont {Miao}\ \emph {et~al.}(2013)\citenamefont {Miao},
  \citenamefont {Huang}, \citenamefont {Qu},\ and\ \citenamefont
  {Chien}}]{FSHEExpMiao}%
  \BibitemOpen
  \bibfield  {author} {\bibinfo {author} {\bibfnamefont {B.~F.}\ \bibnamefont
  {Miao}}, \bibinfo {author} {\bibfnamefont {S.~Y.}\ \bibnamefont {Huang}},
  \bibinfo {author} {\bibfnamefont {D.}~\bibnamefont {Qu}}, \ and\ \bibinfo
  {author} {\bibfnamefont {C.~L.}\ \bibnamefont {Chien}},\ }\href {\doibase
  10.1103/PhysRevLett.111.066602} {\bibfield  {journal} {\bibinfo  {journal}
  {Phys. Rev. Lett.}\ }\textbf {\bibinfo {volume} {111}},\ \bibinfo {pages}
  {066602} (\bibinfo {year} {2013})}\BibitemShut {NoStop}%
\bibitem [{\citenamefont {Wu}\ \emph {et~al.}(2017)\citenamefont {Wu},
  \citenamefont {Wang}, \citenamefont {Huang}, \citenamefont {Qin},
  \citenamefont {Fang}, \citenamefont {Zhang}, \citenamefont {Wan},\ and\
  \citenamefont {Han}}]{FSHEExpWu}%
  \BibitemOpen
  \bibfield  {author} {\bibinfo {author} {\bibfnamefont {H.}~\bibnamefont
  {Wu}}, \bibinfo {author} {\bibfnamefont {X.}~\bibnamefont {Wang}}, \bibinfo
  {author} {\bibfnamefont {L.}~\bibnamefont {Huang}}, \bibinfo {author}
  {\bibfnamefont {J.}~\bibnamefont {Qin}}, \bibinfo {author} {\bibfnamefont
  {C.}~\bibnamefont {Fang}}, \bibinfo {author} {\bibfnamefont {X.}~\bibnamefont
  {Zhang}}, \bibinfo {author} {\bibfnamefont {C.}~\bibnamefont {Wan}}, \ and\
  \bibinfo {author} {\bibfnamefont {X.}~\bibnamefont {Han}},\ }\href {\doibase
  https://doi.org/10.1016/j.jmmm.2017.05.031} {\bibfield  {journal} {\bibinfo
  {journal} {Journal of Magnetism and Magnetic Materials}\ }\textbf {\bibinfo
  {volume} {441}},\ \bibinfo {pages} {149 } (\bibinfo {year}
  {2017})}\BibitemShut {NoStop}%
\bibitem [{\citenamefont {Gibbons}\ \emph {et~al.}(2018)\citenamefont
  {Gibbons}, \citenamefont {MacNeill}, \citenamefont {Buhrman},\ and\
  \citenamefont {Ralph}}]{FSHEExpGibbons}%
  \BibitemOpen
  \bibfield  {author} {\bibinfo {author} {\bibfnamefont {J.~D.}\ \bibnamefont
  {Gibbons}}, \bibinfo {author} {\bibfnamefont {D.}~\bibnamefont {MacNeill}},
  \bibinfo {author} {\bibfnamefont {R.~A.}\ \bibnamefont {Buhrman}}, \ and\
  \bibinfo {author} {\bibfnamefont {D.~C.}\ \bibnamefont {Ralph}},\ }\href
  {\doibase 10.1103/PhysRevApplied.9.064033} {\bibfield  {journal} {\bibinfo
  {journal} {Phys. Rev. Applied}\ }\textbf {\bibinfo {volume} {9}},\ \bibinfo
  {pages} {064033} (\bibinfo {year} {2018})}\BibitemShut {NoStop}%
\bibitem [{\citenamefont {Bose}\ \emph {et~al.}(2018)\citenamefont {Bose},
  \citenamefont {Lam}, \citenamefont {Bhuktare}, \citenamefont {Dutta},
  \citenamefont {Singh}, \citenamefont {Jibiki}, \citenamefont {Goto},
  \citenamefont {Miwa},\ and\ \citenamefont {Tulapurkar}}]{FSHEExpBose}%
  \BibitemOpen
  \bibfield  {author} {\bibinfo {author} {\bibfnamefont {A.}~\bibnamefont
  {Bose}}, \bibinfo {author} {\bibfnamefont {D.~D.}\ \bibnamefont {Lam}},
  \bibinfo {author} {\bibfnamefont {S.}~\bibnamefont {Bhuktare}}, \bibinfo
  {author} {\bibfnamefont {S.}~\bibnamefont {Dutta}}, \bibinfo {author}
  {\bibfnamefont {H.}~\bibnamefont {Singh}}, \bibinfo {author} {\bibfnamefont
  {Y.}~\bibnamefont {Jibiki}}, \bibinfo {author} {\bibfnamefont
  {M.}~\bibnamefont {Goto}}, \bibinfo {author} {\bibfnamefont {S.}~\bibnamefont
  {Miwa}}, \ and\ \bibinfo {author} {\bibfnamefont {A.~A.}\ \bibnamefont
  {Tulapurkar}},\ }\href {\doibase 10.1103/PhysRevApplied.9.064026} {\bibfield
  {journal} {\bibinfo  {journal} {Phys. Rev. Applied}\ }\textbf {\bibinfo
  {volume} {9}},\ \bibinfo {pages} {064026} (\bibinfo {year}
  {2018})}\BibitemShut {NoStop}%
\bibitem [{\citenamefont {Das}\ \emph {et~al.}(2017)\citenamefont {Das},
  \citenamefont {Schoemaker}, \citenamefont {van Wees},\ and\ \citenamefont
  {Vera-Marun}}]{FSHEExpDas}%
  \BibitemOpen
  \bibfield  {author} {\bibinfo {author} {\bibfnamefont {K.~S.}\ \bibnamefont
  {Das}}, \bibinfo {author} {\bibfnamefont {W.~Y.}\ \bibnamefont {Schoemaker}},
  \bibinfo {author} {\bibfnamefont {B.~J.}\ \bibnamefont {van Wees}}, \ and\
  \bibinfo {author} {\bibfnamefont {I.~J.}\ \bibnamefont {Vera-Marun}},\ }\href
  {\doibase 10.1103/PhysRevB.96.220408} {\bibfield  {journal} {\bibinfo
  {journal} {Phys. Rev. B}\ }\textbf {\bibinfo {volume} {96}},\ \bibinfo
  {pages} {220408} (\bibinfo {year} {2017})}\BibitemShut {NoStop}%
\bibitem [{\citenamefont {Baek}\ \emph {et~al.}(2018)\citenamefont {Baek},
  \citenamefont {Amin}, \citenamefont {Oh}, \citenamefont {Go}, \citenamefont
  {Lee}, \citenamefont {Stiles}, \citenamefont {Park},\ and\ \citenamefont
  {Lee}}]{iSOCBaekAmin}%
  \BibitemOpen
  \bibfield  {author} {\bibinfo {author} {\bibfnamefont {S.~C.}\ \bibnamefont
  {Baek}}, \bibinfo {author} {\bibfnamefont {V.~P.}\ \bibnamefont {Amin}},
  \bibinfo {author} {\bibfnamefont {Y.}~\bibnamefont {Oh}}, \bibinfo {author}
  {\bibfnamefont {G.}~\bibnamefont {Go}}, \bibinfo {author} {\bibfnamefont
  {S.-J.}\ \bibnamefont {Lee}}, \bibinfo {author} {\bibfnamefont {M.~D.}\
  \bibnamefont {Stiles}}, \bibinfo {author} {\bibfnamefont {B.-G.}\
  \bibnamefont {Park}}, \ and\ \bibinfo {author} {\bibfnamefont {K.-J.}\
  \bibnamefont {Lee}},\ }\href {\doibase 10.1038/s41563-018-0041-5ID}
  {\bibfield  {journal} {\bibinfo  {journal} {Nature Materials}\ }\textbf
  {\bibinfo {volume} {17}},\ \bibinfo {pages} {509} (\bibinfo {year}
  {2018})}\BibitemShut {NoStop}%
\bibitem [{\citenamefont {Tian}\ \emph {et~al.}(2016)\citenamefont {Tian},
  \citenamefont {Li}, \citenamefont {Qu}, \citenamefont {Huang}, \citenamefont
  {Jin},\ and\ \citenamefont {Chien}}]{FSHEExpTian}%
  \BibitemOpen
  \bibfield  {author} {\bibinfo {author} {\bibfnamefont {D.}~\bibnamefont
  {Tian}}, \bibinfo {author} {\bibfnamefont {Y.}~\bibnamefont {Li}}, \bibinfo
  {author} {\bibfnamefont {D.}~\bibnamefont {Qu}}, \bibinfo {author}
  {\bibfnamefont {S.~Y.}\ \bibnamefont {Huang}}, \bibinfo {author}
  {\bibfnamefont {X.}~\bibnamefont {Jin}}, \ and\ \bibinfo {author}
  {\bibfnamefont {C.~L.}\ \bibnamefont {Chien}},\ }\href {\doibase
  10.1103/PhysRevB.94.020403} {\bibfield  {journal} {\bibinfo  {journal} {Phys.
  Rev. B}\ }\textbf {\bibinfo {volume} {94}},\ \bibinfo {pages} {020403}
  (\bibinfo {year} {2016})}\BibitemShut {NoStop}%
\bibitem [{\citenamefont {Omori}\ \emph {et~al.}(2019)\citenamefont {Omori},
  \citenamefont {Sagasta}, \citenamefont {Niimi}, \citenamefont {Gradhand},
  \citenamefont {Hueso}, \citenamefont {Casanova},\ and\ \citenamefont
  {Otani}}]{FSHEOmori}%
  \BibitemOpen
  \bibfield  {author} {\bibinfo {author} {\bibfnamefont {Y.}~\bibnamefont
  {Omori}}, \bibinfo {author} {\bibfnamefont {E.}~\bibnamefont {Sagasta}},
  \bibinfo {author} {\bibfnamefont {Y.}~\bibnamefont {Niimi}}, \bibinfo
  {author} {\bibfnamefont {M.}~\bibnamefont {Gradhand}}, \bibinfo {author}
  {\bibfnamefont {L.~E.}\ \bibnamefont {Hueso}}, \bibinfo {author}
  {\bibfnamefont {F.}~\bibnamefont {Casanova}}, \ and\ \bibinfo {author}
  {\bibfnamefont {Y.}~\bibnamefont {Otani}},\ }\href {\doibase
  10.1103/PhysRevB.99.014403} {\bibfield  {journal} {\bibinfo  {journal} {Phys.
  Rev. B}\ }\textbf {\bibinfo {volume} {99}},\ \bibinfo {pages} {014403}
  (\bibinfo {year} {2019})}\BibitemShut {NoStop}%
\bibitem [{\citenamefont {Haney}\ and\ \citenamefont
  {Stiles}(2010)}]{Haney:2010}%
  \BibitemOpen
  \bibfield  {author} {\bibinfo {author} {\bibfnamefont {P.~M.}\ \bibnamefont
  {Haney}}\ and\ \bibinfo {author} {\bibfnamefont {M.~D.}\ \bibnamefont
  {Stiles}},\ }\href {\doibase 10.1103/PhysRevLett.105.126602} {\bibfield
  {journal} {\bibinfo  {journal} {Phys. Rev. Lett.}\ }\textbf {\bibinfo
  {volume} {105}},\ \bibinfo {pages} {126602} (\bibinfo {year}
  {2010})}\BibitemShut {NoStop}%
\bibitem [{\citenamefont {Kundt}(1893)}]{AHETheoryKundt}%
  \BibitemOpen
  \bibfield  {author} {\bibinfo {author} {\bibfnamefont {A.}~\bibnamefont
  {Kundt}},\ }\href {\doibase 10.1002/andp.18932850603} {\bibfield  {journal}
  {\bibinfo  {journal} {Annalen der Physik}\ }\textbf {\bibinfo {volume}
  {285}},\ \bibinfo {pages} {257} (\bibinfo {year} {1893})}\BibitemShut
  {NoStop}%
\bibitem [{\citenamefont {Pugh}\ and\ \citenamefont
  {Rostoker}(1953)}]{AHETheoryPughRostoker}%
  \BibitemOpen
  \bibfield  {author} {\bibinfo {author} {\bibfnamefont {E.~M.}\ \bibnamefont
  {Pugh}}\ and\ \bibinfo {author} {\bibfnamefont {N.}~\bibnamefont
  {Rostoker}},\ }\href {\doibase 10.1103/RevModPhys.25.151} {\bibfield
  {journal} {\bibinfo  {journal} {Rev. Mod. Phys.}\ }\textbf {\bibinfo {volume}
  {25}},\ \bibinfo {pages} {151} (\bibinfo {year} {1953})}\BibitemShut
  {NoStop}%
\bibitem [{\citenamefont {Karplus}\ and\ \citenamefont
  {Luttinger}(1954)}]{AHETheoryKarplusLuttinger}%
  \BibitemOpen
  \bibfield  {author} {\bibinfo {author} {\bibfnamefont {R.}~\bibnamefont
  {Karplus}}\ and\ \bibinfo {author} {\bibfnamefont {J.~M.}\ \bibnamefont
  {Luttinger}},\ }\href {\doibase 10.1103/PhysRev.95.1154} {\bibfield
  {journal} {\bibinfo  {journal} {Phys. Rev.}\ }\textbf {\bibinfo {volume}
  {95}},\ \bibinfo {pages} {1154} (\bibinfo {year} {1954})}\BibitemShut
  {NoStop}%
\bibitem [{\citenamefont {Nagaosa}\ \emph {et~al.}(2010)\citenamefont
  {Nagaosa}, \citenamefont {Sinova}, \citenamefont {Onoda}, \citenamefont
  {MacDonald},\ and\ \citenamefont {Ong}}]{AHEReviewNagosa}%
  \BibitemOpen
  \bibfield  {author} {\bibinfo {author} {\bibfnamefont {N.}~\bibnamefont
  {Nagaosa}}, \bibinfo {author} {\bibfnamefont {J.}~\bibnamefont {Sinova}},
  \bibinfo {author} {\bibfnamefont {S.}~\bibnamefont {Onoda}}, \bibinfo
  {author} {\bibfnamefont {A.~H.}\ \bibnamefont {MacDonald}}, \ and\ \bibinfo
  {author} {\bibfnamefont {N.~P.}\ \bibnamefont {Ong}},\ }\href {\doibase
  10.1103/RevModPhys.82.1539} {\bibfield  {journal} {\bibinfo  {journal} {Rev.
  Mod. Phys.}\ }\textbf {\bibinfo {volume} {82}},\ \bibinfo {pages} {1539}
  (\bibinfo {year} {2010})}\BibitemShut {NoStop}%
\bibitem [{\citenamefont {Taniguchi}\ \emph {et~al.}(2015)\citenamefont
  {Taniguchi}, \citenamefont {Grollier},\ and\ \citenamefont
  {Stiles}}]{AHEAMRTaniguchi}%
  \BibitemOpen
  \bibfield  {author} {\bibinfo {author} {\bibfnamefont {T.}~\bibnamefont
  {Taniguchi}}, \bibinfo {author} {\bibfnamefont {J.}~\bibnamefont {Grollier}},
  \ and\ \bibinfo {author} {\bibfnamefont {M.~D.}\ \bibnamefont {Stiles}},\
  }\href {\doibase 10.1103/PhysRevApplied.3.044001} {\bibfield  {journal}
  {\bibinfo  {journal} {Phys. Rev. Applied}\ }\textbf {\bibinfo {volume} {3}},\
  \bibinfo {pages} {044001} (\bibinfo {year} {2015})}\BibitemShut {NoStop}%
\bibitem [{\citenamefont {Giannozzi}\ \emph {et~al.}(2009)\citenamefont
  {Giannozzi}, \citenamefont {Baroni}, \citenamefont {Bonini}, \citenamefont
  {Calandra}, \citenamefont {Car}, \citenamefont {Cavazzoni}, \citenamefont
  {Ceresoli}, \citenamefont {Chiarotti}, \citenamefont {Cococcioni},
  \citenamefont {Dabo} \emph {et~al.}}]{giannozzi2009quantum}%
  \BibitemOpen
  \bibfield  {author} {\bibinfo {author} {\bibfnamefont {P.}~\bibnamefont
  {Giannozzi}}, \bibinfo {author} {\bibfnamefont {S.}~\bibnamefont {Baroni}},
  \bibinfo {author} {\bibfnamefont {N.}~\bibnamefont {Bonini}}, \bibinfo
  {author} {\bibfnamefont {M.}~\bibnamefont {Calandra}}, \bibinfo {author}
  {\bibfnamefont {R.}~\bibnamefont {Car}}, \bibinfo {author} {\bibfnamefont
  {C.}~\bibnamefont {Cavazzoni}}, \bibinfo {author} {\bibfnamefont
  {D.}~\bibnamefont {Ceresoli}}, \bibinfo {author} {\bibfnamefont {G.~L.}\
  \bibnamefont {Chiarotti}}, \bibinfo {author} {\bibfnamefont {M.}~\bibnamefont
  {Cococcioni}}, \bibinfo {author} {\bibfnamefont {I.}~\bibnamefont {Dabo}},
  \emph {et~al.},\ }\href@noop {} {\bibfield  {journal} {\bibinfo  {journal}
  {Journal of physics: Condensed matter}\ }\textbf {\bibinfo {volume} {21}},\
  \bibinfo {pages} {395502} (\bibinfo {year} {2009})}\BibitemShut {NoStop}%
\bibitem [{\citenamefont {Fuh}\ and\ \citenamefont
  {Guo}(2011)}]{fuh2011intrinsic}%
  \BibitemOpen
  \bibfield  {author} {\bibinfo {author} {\bibfnamefont {H.-R.}\ \bibnamefont
  {Fuh}}\ and\ \bibinfo {author} {\bibfnamefont {G.-Y.}\ \bibnamefont {Guo}},\
  }\href@noop {} {\bibfield  {journal} {\bibinfo  {journal} {Physical Review
  B}\ }\textbf {\bibinfo {volume} {84}},\ \bibinfo {pages} {144427} (\bibinfo
  {year} {2011})}\BibitemShut {NoStop}%
\bibitem [{\citenamefont {Mostofi}\ \emph {et~al.}(2008)\citenamefont
  {Mostofi}, \citenamefont {Yates}, \citenamefont {Lee}, \citenamefont {Souza},
  \citenamefont {Vanderbilt},\ and\ \citenamefont
  {Marzari}}]{mostofi2008wannier90}%
  \BibitemOpen
  \bibfield  {author} {\bibinfo {author} {\bibfnamefont {A.~A.}\ \bibnamefont
  {Mostofi}}, \bibinfo {author} {\bibfnamefont {J.~R.}\ \bibnamefont {Yates}},
  \bibinfo {author} {\bibfnamefont {Y.-S.}\ \bibnamefont {Lee}}, \bibinfo
  {author} {\bibfnamefont {I.}~\bibnamefont {Souza}}, \bibinfo {author}
  {\bibfnamefont {D.}~\bibnamefont {Vanderbilt}}, \ and\ \bibinfo {author}
  {\bibfnamefont {N.}~\bibnamefont {Marzari}},\ }\href@noop {} {\bibfield
  {journal} {\bibinfo  {journal} {Computer physics communications}\ }\textbf
  {\bibinfo {volume} {178}},\ \bibinfo {pages} {685} (\bibinfo {year}
  {2008})}\BibitemShut {NoStop}%
\bibitem [{Note1()}]{Note1}%
  \BibitemOpen
  \bibinfo {note} {We find small deviations from the symmetry-required form of
  the conductivity. This is due to the fact that the Wannier projection results
  in a Hamiltonian which does not exactly respect the symmetries of the crystal
  lattice. We have also constructed symmetrized Hamiltonians, and find that
  symmetry-related errors introduced by the Wannier process are small, on the
  order of $(0.01~,0.1,~10)~\left ({\protect \rm \Omega \cdot cm}\right )^{-1}$
  for Fe, Ni, and Co, respectively.}\BibitemShut {Stop}%
\bibitem [{Note2()}]{Note2}%
  \BibitemOpen
  \bibinfo {note} {The general form for the spin current tensor for flow in the
  ${\protect \bf {\protect \mathaccentV {hat}05En}}$-direction, spin-polarized
  in the ${\protect \bf {\protect \mathaccentV {hat}05Es}}$-direction under an
  applied field ${\protect \bf E}$ is: $Q^\protect \mathbf {\protect
  \mathaccentV {hat}05E{n}}_\protect \mathbf {\protect \mathaccentV
  {hat}05E{s}} = \sigma _{\parallel }\left ({{\protect \bf \protect
  \mathaccentV {hat}05Es}} \cdot {{\protect \bf \protect \mathaccentV
  {hat}05Em}}\right ) | {{\protect \bf \protect \mathaccentV {hat}05Em}} \cdot
  \left ({\protect \bf E}\times {{\protect \bf \protect \mathaccentV
  {hat}05En}}\right )| + \sigma _{\perp } {{\protect \bf \protect \mathaccentV
  {hat}05Es}} \cdot \left ( {{\protect \bf \protect \mathaccentV {hat}05Em}}
  \times \left ( {{\protect \bf \protect \mathaccentV {hat}05Em}} \times \left
  ( {\protect \bf E} \times {{\protect \bf \protect \mathaccentV {hat}05En}}
  \right ) \right ) \right )$}\BibitemShut {NoStop}%
\bibitem [{Note3()}]{Note3}%
  \BibitemOpen
  \bibinfo {note} {Ref. \cite {wang2006ab} partitions contributions to $\sigma
  _{\protect \rm AHE}$ according to band-pair type and finds a different
  values. However the sum is partitioned differently in the two cases: Ref.
  \cite {wang2006ab} work assigns one band-pair type to each ${\protect \bf
  k}$-point, whereas we track all band-pair types.}\BibitemShut {Stop}%
\bibitem [{\citenamefont {Tanaka}\ \emph {et~al.}(2008)\citenamefont {Tanaka},
  \citenamefont {Kontani}, \citenamefont {Naito}, \citenamefont {Naito},
  \citenamefont {Hirashima}, \citenamefont {Yamada},\ and\ \citenamefont
  {Inoue}}]{SHETheoryTanaka}%
  \BibitemOpen
  \bibfield  {author} {\bibinfo {author} {\bibfnamefont {T.}~\bibnamefont
  {Tanaka}}, \bibinfo {author} {\bibfnamefont {H.}~\bibnamefont {Kontani}},
  \bibinfo {author} {\bibfnamefont {M.}~\bibnamefont {Naito}}, \bibinfo
  {author} {\bibfnamefont {T.}~\bibnamefont {Naito}}, \bibinfo {author}
  {\bibfnamefont {D.~S.}\ \bibnamefont {Hirashima}}, \bibinfo {author}
  {\bibfnamefont {K.}~\bibnamefont {Yamada}}, \ and\ \bibinfo {author}
  {\bibfnamefont {J.}~\bibnamefont {Inoue}},\ }\href {\doibase
  10.1103/PhysRevB.77.165117} {\bibfield  {journal} {\bibinfo  {journal} {Phys.
  Rev. B}\ }\textbf {\bibinfo {volume} {77}},\ \bibinfo {pages} {165117}
  (\bibinfo {year} {2008})}\BibitemShut {NoStop}%
\bibitem [{\citenamefont {Kontani}\ \emph {et~al.}(2007)\citenamefont
  {Kontani}, \citenamefont {Tanaka},\ and\ \citenamefont
  {Yamada}}]{kontani2007intrinsic}%
  \BibitemOpen
  \bibfield  {author} {\bibinfo {author} {\bibfnamefont {H.}~\bibnamefont
  {Kontani}}, \bibinfo {author} {\bibfnamefont {T.}~\bibnamefont {Tanaka}}, \
  and\ \bibinfo {author} {\bibfnamefont {K.}~\bibnamefont {Yamada}},\
  }\href@noop {} {\bibfield  {journal} {\bibinfo  {journal} {Physical Review
  B}\ }\textbf {\bibinfo {volume} {75}},\ \bibinfo {pages} {184416} (\bibinfo
  {year} {2007})}\BibitemShut {NoStop}%
\bibitem [{\citenamefont {Park}\ \emph {et~al.}(2011)\citenamefont {Park},
  \citenamefont {Kim}, \citenamefont {Yu}, \citenamefont {Han},\ and\
  \citenamefont {Kim}}]{RasbhaTheoryPark}%
  \BibitemOpen
  \bibfield  {author} {\bibinfo {author} {\bibfnamefont {S.~R.}\ \bibnamefont
  {Park}}, \bibinfo {author} {\bibfnamefont {C.~H.}\ \bibnamefont {Kim}},
  \bibinfo {author} {\bibfnamefont {J.}~\bibnamefont {Yu}}, \bibinfo {author}
  {\bibfnamefont {J.~H.}\ \bibnamefont {Han}}, \ and\ \bibinfo {author}
  {\bibfnamefont {C.}~\bibnamefont {Kim}},\ }\href {\doibase
  10.1103/PhysRevLett.107.156803} {\bibfield  {journal} {\bibinfo  {journal}
  {Phys. Rev. Lett.}\ }\textbf {\bibinfo {volume} {107}},\ \bibinfo {pages}
  {156803} (\bibinfo {year} {2011})}\BibitemShut {NoStop}%
\bibitem [{\citenamefont {Wang}\ \emph {et~al.}(2006)\citenamefont {Wang},
  \citenamefont {Yates}, \citenamefont {Souza},\ and\ \citenamefont
  {Vanderbilt}}]{wang2006ab}%
  \BibitemOpen
  \bibfield  {author} {\bibinfo {author} {\bibfnamefont {X.}~\bibnamefont
  {Wang}}, \bibinfo {author} {\bibfnamefont {J.~R.}\ \bibnamefont {Yates}},
  \bibinfo {author} {\bibfnamefont {I.}~\bibnamefont {Souza}}, \ and\ \bibinfo
  {author} {\bibfnamefont {D.}~\bibnamefont {Vanderbilt}},\ }\href@noop {}
  {\bibfield  {journal} {\bibinfo  {journal} {Physical Review B}\ }\textbf
  {\bibinfo {volume} {74}},\ \bibinfo {pages} {195118} (\bibinfo {year}
  {2006})}\BibitemShut {NoStop}%
\end{thebibliography}%


\end{document}